\documentclass[twocolumn,showpacs,prb,amsmath,amssymb]{revtex4}
\usepackage{graphicx}
\usepackage{dcolumn}
\usepackage{bm}

\begin{document}

\title{Current-voltage characteristics of graphene devices: interplay between Zener-Klein tunneling and defects}
\author{Niels Vandecasteele$^1$, Amelia Barreiro$^2$, Michele Lazzeri$^1$,
Adrian Bachtold$^2$, and Francesco Mauri$^1$}
\affiliation{$^1$ IMPMC, Universit\'es Paris 6 et 7, CNRS, IPGP,
140 rue de Lourmel, 75015 Paris, France\\
$^2$ CIN2(CSIC-ICN), Campus UAB, E-08193 Barcelona, Spain}

\begin{abstract}
We report a theoretical/experimental study of current-voltage
characteristics ($I$-$V$) of graphene devices near the Dirac point.
The $I$-$V$ can be described by a power law ($I\propto V^\alpha$, with
$1<\alpha\le1.5$).  The exponent is higher when the mobility is lower.
This superlinear $I$-$V$ is interpreted in terms of the interplay
between Zener-Klein transport, that is tunneling between different energy
bands, and defect scattering.  Surprisingly, the Zener-Klein tunneling
is made visible by the presence of defects.

\end{abstract}


\maketitle

Zener tunneling~\cite{zener34} is a concept known since the 30's,
which, in a solid, refers to the tunneling of carriers from one band
to another through the forbidden energy gap (for example from the
conduction to the valence band).  This tunnel process is very
intriguing in graphene because the energy gap is suppressed to zero
and because of the peculiar charge carriers behaving as Dirac fermions
~\cite{novoselov04,katsnelson06}.  In
particular, some of the carriers (those with the velocity parallel to
the electric field) experience Zener tunneling without being
backscattered
~\cite{katsnelson06,cheianov06,fogler08,sonin09}, a behavior
which is markedly different from the one in conventional
semiconductors.  The physics is the same as for relativistic electrons
tunneling through a barrier, a phenomenon called Klein
tunneling~\cite{klein29} and, for this reason, we will use the term
Zener-Klein (ZK) tunneling.

In view of the remarkable properties of ZK tunneling in graphene, it
is understandable that an intensive endeavor was made to challenge it.
So far, the effort was focused on graphene $p$-$n$
junctions~\cite{lemme07,huard07,gorbachev08,liu08,stander09,young09}.
In these devices, carriers tunnel through a sharp energy barrier
induced with external local gate electrodes.  Sophisticated
nanofabrication techniques were employed to structure these local
gates. For instance the insulator layer was very
thin~\cite{lemme07,huard07,stander09}, the local gate was separated
from the graphene by an air gap ~\cite{gorbachev08,liu08}, or the
local gate was extremely narrow~\cite{ young09}.

Here, we argue that Zener-Klein tunneling can be observed
in graphene with the most common device layout
(undoped, four-point configuration, and without any local gates)
by simply measuring the $I$-$V$ at room temperature.
First, we provide  an analytical semi-classical expression for the $I$-$V$s
as a function of the doping. In graphene, the ZK current manifests itself
with a superlinear current $I\propto V^{\alpha}$, with $\alpha=1.5$.
Then, we study the role of defects with the ``exact'' (non-perturbative)
non-equilibrium Green-function approach finding the counterintuitive
result that charged impurities enhance the visibility of the ZK current.
Finally, we report measurements  showing
that the $I$-$V$s at the Dirac point is indeed described by power
laws, $I\propto V^\alpha$, with $\alpha$ ranging from 1 to 1.4.
The exponent $\alpha$ is higher when the mobility is lower, consistently
with our theoretical predictions.

\begin{figure}[h!]
\centerline{\includegraphics[width=85mm]{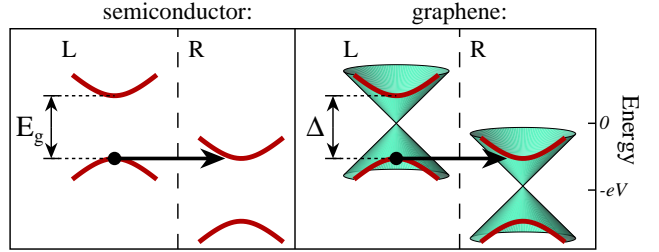}}
\caption{(Color online) Bands of the L and R contacts
in a semiconductor and in graphene. The arrows represent the possible occurrence
of Zener-Klein tunneling.}
\label{fig1}
\end{figure}

In graphene ZK tunneling leads to unusual $I$-$V$s as compared to
those of semiconductors/insulators.
Let us consider transport through a piece of a material and apply a voltage
-$V$ between the right (R) and left (L) sides.
For a semiconductor with electronic gap ${\rm E_g}$, ZK
tunneling is possible only for $eV>{\rm E_g}$, where $e>0$ is the electron charge
(Fig.~\ref{fig1}).
On the contrary, in graphene (usually defined as a semi-metal)
the gap is zero and, thus, ZK tunneling is possible for arbitrarily small $V$.

More specifically, the two-dimensional electronic-band dispersion of graphene is a cone:
$\epsilon=\pm\hbar v_F\sqrt{k_\perp^2+k^2}$, where $k$ ($k_\perp$) is the
wavevector-component parallel (perpendicular) to the current flow.
During ballistic transport (in absence of scattering) $k_\perp$ is conserved.
For a fixed $k_\perp$, the bands are hyperbolae with gap
$\Delta=2\hbar v_F k_\perp$ (Fig.~\ref{fig1}).
For any $V$, there are conducting channels for which the tunneling
is possible (with $k_\perp$ such that $\Delta<V$).
We will show that this results in a tunneling current $I\propto V^{3/2}$.
By contrast, the ZK tunneling current in semiconductors
vanishes exponentially at low $V$.

In graphene, 
within the Landauer approach, the current per unit of lateral length, J, is
\begin{eqnarray}
J& =& \frac{4e}{h} \int \frac{dk_\perp}{2\pi} \int_{\epsilon_F-eV}^{\epsilon_F}
T(\epsilon,k_\perp,V) d\epsilon = \nonumber \\
&=& \frac{4e}{h} \int_{\epsilon_F-eV}^{\epsilon_F} {\cal T}(\epsilon,V) d\epsilon
\label{eq1}
\end{eqnarray}
where the factor 4 accounts for spin and valley degeneracy
and the transmission $T(\epsilon,k_\perp,V)$
is the probability that an electron
(with energy $\epsilon$ and perpendicular momentum $k_\perp$)
is transmitted through the channel.
We assume a uniform drop of the electrostatic potential along the 
current-flow direction,
with constant electric field  $V/l$, being $l$ the distance between the
contacts.

\begin{figure}[h!]
\centerline{\includegraphics[width=85mm]{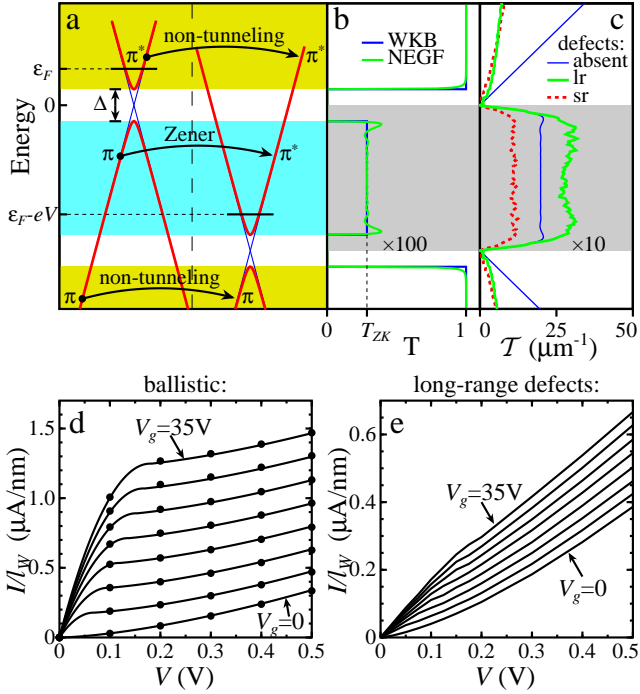}}
\caption{(Color online)
Electronic transport in graphene: theory.
a: electronic bands of the left (L)
and right (R) contacts.
The hyperbolae are the bands corresponding to a finite $k_\perp$
($\Delta=2\hbar v_Fk_\perp=22$~meV).
In the L and R contacts the bands are filled up to $\epsilon_F$ and $\epsilon_F-eV$,
respectively, where $\epsilon_F>0$ ($\epsilon_F<0$) corresponds to electron (hole)
doping.
b,c: electronic transmissions $T$ and ${\cal T}$, defined as in Eq.~\ref{eq1},
for $V=0.1$~V and $l=1~\mu$m. In the gray zone (the region corresponding to the
Zener-Klein tunneling) $T$ and ${\cal T}$ are magnified.
b: ballistic case calculated with NEGF or with the WKB approximation.
c: NEGF results in the ballistic case (no defects) or in the presence of
long-range (lr) or short-range (sr) defects.
d,e: calculated current $I$ per unit of lateral-length $l_W$
vs. $V$ (the voltage applied between the electrodes)
as a function of the gate voltage ($V_g$).
$V_g$ goes from 0 to 35~V with 5~V step, $l=1~\mu$m.
In the ballistic case (d), lines are approximated semiclassical results
(i.e. the analytical curves from note~\onlinecite{iv_anal}),
points are from the ``exact'' NEGF simulations.
}
\label{fig2}
\end{figure}

\begin{figure}[h!]
\centerline{\includegraphics[width=85mm]{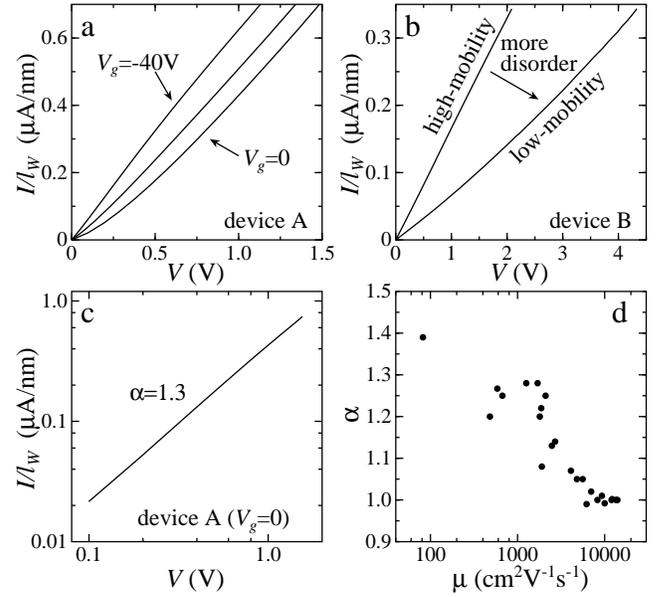}}
\caption{Electronic transport in graphene: measurements.
a: measured current $I$ per unit of lateral-length $l_W$ vs. $V$ in a low-mobility sample,
for different gate values ($V_g$=0,-20,-40~V).
$V_g$ has been shifted to -14~V so that $V_g$=0 corresponds
to the Dirac point.
The length between the voltage electrodes $l=1.1~\mu$m and $l_W=1.1~\mu$m.
b: measured $I$-$V$ in a high-mobility sample, at the Dirac point,
before and after the introduction of defects through electronic bombardment.
$l$=2.2~$\mu$m and $l_W$=550~nm.
c: double-logarithmic scale plot of the $I$-$V$. 
d: exponent $\alpha$ as a function of mobility $\mu$ for different devices.
$l$ varies from 0.9 to 5.9~$\mu$m and $l_W$ from 70 to 1500~nm.
}
\label{fig3}
\end{figure}

The transmission can be calculated with the non equilibrium Green function
(NEGF) method. To describe the purely ballistic case, we also use a 
semiclassical approach, based on the Wentzel-Kramers-Brillouin (WKB) approximation.
The transmission $T^{WKB}(\epsilon,k_\perp,V)$ can be equal to 1, 0, or to
$T_{ZK}=\exp[-\pi l \Delta^2 / (4\hbar v_F e V)]$~\cite{cheianov06,andreev07,sonin09}
(see the example in Fig.~\ref{fig2}ab).
We call non-tunneling current (Fig.~\ref{fig2}a), the one associated with carriers that 
always remain in the same band $\pi$ or $\pi^*$
($T^{WKB}=1$, light-shadowed (yelow) area in Fig.~\ref{fig2}a).
We call ``Zener-Klein'' current, the one associated with carriers that tunnel
from the $\pi$ to the $\pi^*$ band ($T^{WKB}=T_{ZK}$, dark-shadowed (cyan) area in Fig.~\ref{fig2}a).
From Fig.~\ref{fig2}b,  the $T^{WKB}$ transmission is a good approximation to
our most precise NEGF calculations~\cite{NEGF}.


In a graphene-based field-effect device, the density of the carriers
$n$ can be varied by changing the gate voltage $V_g$.  $n=V_gC_g/e$,
being $C_g$ the gate-channel capacitance.  Fig.~\ref{fig2}d reports
the current-voltage ($I$-$V$) curves in the ballistic regime obtained
with the semiclassical WKB approach (by letting $T=T^{WKB}$ in Eq.~\ref{eq1})
and with the ``exact'' NEGF
method~\cite{NEGF}, for various dopings (we use
$C_g=1.15\times10^{-4}$~F/m$^2$~~\cite{barreiro09}). 
The two methods give almost identical results
(for the WKB case, we report in note~\onlinecite{iv_anal} an analytical expression
for the $I$-$V$ as a function of $\epsilon_F$).
For zero-doping ($V_g=0$~V) there is no
contribution from the non-tunneling current, the current is entirely
due to ZK tunneling, and the $I$-$V$ curve is superlinear
($I\propto V^{3/2}$)~\cite{iv_anal}.  As soon as the system is doped
(already for $V_g=5$~V) the ZK current is no more dominant 
(with respect to the non-tunneling current) and for
small bias ($V<0.1$~V) the $I$-$V$ is ohmic (linear).

Do we expect the superlinear ZK current to be visible in actual
devices?  At first sight the answer is no for two reasons.  First, in
actual devices, the carrier concentration is never exactly zero.
Indeed it has been observed~\cite{hwang07} that the presence of charge
impurities induces
a spatial fluctuation of the Fermi level with respect to the Dirac
point.  As a result, it is difficult to achieve the experimental condition
where ZK tunneling is observable ($V_g$=0 in Fig.~\ref{fig2}d).
Second, the scattering of the carriers with optical phonons with
energy $\hbar\omega$=0.15~eV causes the current to saturate when
increasing $V$ to high values~\cite{barreiro09,kim}.
This process occurs for $eV>\hbar\omega
l/l_{el}$ ($l_{el}$ is the carrier elastic scattering length, due to
defects) and is, thus, particularly relevant for high-quality
high-mobility samples (with high $l_{el}$).  This saturation of the
non-tunneling current induces a sublinearity ($I\propto V^\alpha$,
with $\alpha<1$) which tends to cancel the superlinearity ($\alpha>1$)
of the ZK current, further masking it (see~\onlinecite{suppl_info} for
further discussion).

The situation is possibly changed by the presence of defects.
Actual devices are characterized by defects which
scatter electrons elastically (that is conserving the
energy)~\cite{barreiro09}.  Elastic defects can be neutral point
defects or charged Coulomb impurities~\cite{chen08} outside the
graphene plane (usually at a distance $\sim$~1$-$2 nm) ~\cite{tse08}.
Point defects affect the electrostatic potential seen by the carriers
on a length scale smaller than the graphene unitary cell
($short-range$) and, thus, the carriers cannot be described in terms
of electronic bands.  On the other hand, charged impurities modify the
potential uniformly on a length scale much longer than the unit cell
($long-range$) and the electronic bands are still a meaningful
concept.  The ZK current is expected to be more sensitive to
short-range defects than to long-range ones. Indeed, the ZK current
is determined by a transition between two bands whose relative energy
is not affected by long-range defects.  Moreover, long-range defects
are expected to diminish the non-tunneling current.  Overall,
one could wonder whether {\it the presence of long-range defects can
be used to suppress the non-tunneling current and, thus, to make
visible the ZK one}.

To verify this hypothesis, we simulate disordered graphene within NEGF
by considering both long- and short-range elastic
defects~\cite{NEGF}. We remark that the NEGF approach provides an exact
(non-perturbative) atomistic treatment of disorder.
Defects are simulated by changing randomly the 
on-site potential by $V_d=0.1$~eV. This $V_d$ is
a realistic choice since it provides a low-bias conductivity in
reasonable agreement with measurements~\cite{suppl_info}.

From NEGF simulations, the presence of long-range defects diminishes
the non-tunneling transmission (Fig.~\ref{fig2}c)
but, in general does not reduce the ZK one.
For $V$=0.1~V, long-range defects even increase the ZK
transmission (Fig.~\ref{fig2}c).  
We checked that short-range defects diminish, as
expected, both the non-tunneling and the ZK transmission, with respect to
the ballistic case (Fig.~\ref{fig2}c).  To see whether the
relative increase of the ZK transmission with respect to the non-tunneling
one can lead to measurable
effects, in Fig.~\ref{fig2}e we show the theoretical $I$-$V$ curves in the
presence of long-range defects. The superlinear behavior
(the signature of the ZK current) is still visible at $V_g=0$
($I\propto V^\alpha$, with $\alpha=1.4$ in Fig.~\ref{fig2}e)
and is also visible at finite $V_g$.


We now turn our attention to measurements, carried out on
single-layer graphene devices~\cite{experimental}.
Different devices were fabricated in a four-point configuration
and have different mobilities $\mu$ ranging from
80 to 20000 cm$^2$V$^{-1}$s$^{-1}$ (low mobility corresponds
to a higher density of defects)~\cite{experimental}.
Fig.~\ref{fig3}a shows a typical set of $I$-$V$
characteristics for different $V_g$ applied on the backgate
for a sample with a relatively modest mobility
($\mu=1700$~cm$^2$V$^{-1}$s$^{-1}$).
The $I$-$V$ is superlinear at the $V_g$ of the Dirac point, consistent with the
above prediction of ZK tunneling.  The superlinearity
is better seen in a double-logarithmic scale plot
(Fig.~\ref{fig3}c) where the $I$-$V$ is reasonably well described
by a power law $I\propto V^\alpha$ with $\alpha=1.3$.
Both $\alpha$ and the current values
are in a remarkable qualitative agreement with calculations
given the simplicity of the model as can be seen by comparing
Fig.~\ref{fig2}e ($l$=1$\mu$m) and Fig.~\ref{fig3}a ($l$=1.1$\mu$m) 
for small $V_g$.
More elaborated models ({\it e.g.} with a more realistic description of 
impurities and including electron-phonon scattering) are required
to reach a quantitative agreement between theory and measurements.

We observe that the superlinearity vanishes for devices with high
$\mu$.  Fig.~\ref{fig3}d shows $\alpha$ (extracted at the Dirac point)
as a function of the mobility $\mu$ of 22 different devices. Indeed,
as the mobility increases, $\alpha$ tends to 1 (corresponding to
linear $I$-$V$).  In an additional experiment, we introduced defects
in a high-mobility graphene device by bombarding it with 10 keV
electrons.  From Fig.~\ref{fig3}b, before bombardment the mobility
$\mu=$7000 cm$^2$V$^{-1}$s$^{-1}$ and the $I$-$V$ is linear with
$\alpha=1.0$. After bombardment $\mu$ drops to 260
cm$^2$V$^{-1}$s$^{-1}$ and the $I$-$V$ becomes superlinear
($\alpha=1.2$).  These observations are consistent with the above
discussion that in high-mobility samples the ZK superlinearity is
masked by the non-tunneling current.  Namely, the reduction of
disorder increases the contribution of the non-tunneling current with
respect to the ZK one and, also, favors the non-tunneling
current saturation (due to scattering with optical
phonons~\cite{barreiro09,kim}).

We now discuss other mechanisms that could lead to
superlinear $I$-$V$s. It could be related to the physics occurring at
tunnel barriers (such as the Luttinger liquid-like behavior in
nanotubes or the breakdown of insulating barriers). However,
measurements are done on high-quality devices in a four-point
configuration, which makes the presence of tunnel barriers
unlikely. Superlinear $I$-$V$s could also be attributed to quantum
effects, such as weak localization or electron-electron interaction,
but these effects should be negligible since the applied current is
large, heating the graphene layer to several hundreds of Celsius~\cite{freitag09}.
Overall, Zener-Klein tunneling remains the most plausible mechanism to
explain our measurements.


We finally stress that previous observations of Klein tunneling
~\cite{gorbachev08,young09,stander09}
in graphene were done using very different device setups.
In~\onlinecite{gorbachev08,young09,stander09}, the carriers tunnel from
conduction to valence bands in a $p$-$n$ junction.  In these
nanostructured devices, the ZK tunneling is observed thanks to a
configuration which allows to eliminate the non-tunneling current and
thanks to the intense electric field at the $p$-$n$ junction ($\sim$
10$^{-3}$ eV/\AA, see suppl. info of~\onlinecite{young09} and~\onlinecite{huard07}).
In our devices, which are not $p$-$n$ junctions,
the non-tunneling (intraband) current is present
(this current can mask the ZK tunneling one) and
the electric field ($\sim$10$^{-5}$ eV/\AA) is substantially weaker.
Despite these unfavorable conditions, it is possible to probe the 
Zener-Klein effect.

Concluding, measurements and calculations show, consistently, that the
$I$-$V$s of graphene devices become superlinear in the presence of
disorder (in low-mobility samples).  The superlinearity is attributed
to Zener-Klein tunneling (tunneling between different energy bands,
from $\pi$ to $\pi^*$).  In high-mobility (high-quality)
graphene samples, the superlinearity is masked by the contribution of
the non-tunneling current (due to carriers always remaining the same
band).  In low-mobility samples, the Zener-Klein tunneling current is
visible because the higher density of defects decreases (filters) the
non-tunneling current.

We thank G.Stoltz.
The research was supported by an EURYI Grant, an EU grant No. FP6-
IST-021285-2, the ANR PNANO-ACCATTONE and IDRIS (Orsay).


\begin{thebibliography}{99}
\bibitem{NEGF}
We consider a first-neighbors tight-binding electronic Hamiltonian
with overlap parameter $t=3.0$~eV between $\pi_z$ orbitals
($v_F=10^8$~cm/s) and lattice spacing $a_0=2.46$~\AA.  NEGF equations
are solved as in Ref.~\onlinecite{stoltz09} for a strip with length
$l=1\mu{\rm m}$ and lateral infinite width.  Defect-induced disorder
is simulated by changing the on-site potential by $V_d$=0.1~eV
uniformly on trenches of atoms which are randomly distributed on along
the channel.  $V_d$ is uniform along the direction perpendicular to
the channel (disorder does not break the lateral periodicity) for
computational reasons.  Long (short) range defects are simulated using
trenches long 10$a_0$ ($a_0$).  The results are obtained after
averaging over various disorder configuration.  
See~\onlinecite{suppl_info} for more details.
The present approach
does not include scattering with optical phonons, which is relevant
for $eV>\hbar\omega l/l_{el}$.
\bibitem{iv_anal}
Let us write the total current per unit of lateral length
as $J=J_c+J_Z$, where $J_c$ and $J_Z$ are the non-tunneling
and ZK currents, and let $T=T^{WKB}$ in Eq.~\ref{eq1}.
For $\epsilon_F=0$, $J_c=0$ and $J_Z=2J_0 {\cal F}(\kappa_0)$, where
$J_0=V/(\pi^2R_0l)$, $R_0=h/(2e^2)=12.9$~k$\Omega$, the adimensional
quantity $\kappa_0=V\pi e l/(4\hbar v_F)$, and
${\cal F}(x)=\sqrt{\pi x}~{\rm erf}(\sqrt{x})+e^{-x}-1$.
For $l=1\mu$m, $\kappa_0=V/(0.8~{\rm mV})$ and for $V\gg 0.8$~mV one
can show that $J_Z\simeq2J_0\sqrt{\pi\kappa_0}\propto V^{3/2}$.
For $\epsilon_F\ne0$, let us call $\xi=eV/\epsilon_F$.
For $\xi<1$~~$J_c=J_1(2-\xi)\xi$, while for $\xi>1$ $J_c$ saturates to $J_c=J_1$, where
$J_1=ev_Fn/\pi$, being $n=\epsilon_F^2/\pi/(\hbar v_F)^2$ the electron concentration.
For $\xi<1$~~$J_Z=0$, while
for $1<\xi<2$~~$J_Z=J_0 {\cal F}[\kappa_1(\xi-1)^2]$, and
for $2<\xi$~~$J_Z=J_0 \{2{\cal F}(\kappa_0)-{\cal F}(\kappa_1)\}$,
where the adimensional quantity
$\kappa_1=J_1\pi^2R_0l/V$ ($\kappa_0=\kappa_1\xi^2/4$).
See~\onlinecite{suppl_info} for a derivation.
Note that the calculated saturation of the non-tunneling current at $eV=\epsilon_F$
in the ballistic regime ($J_c=J_1$, see Fig.~\ref{fig2}d)
should not be confused with the saturation due to optical-phonon
scattering of~\onlinecite{barreiro09,kim}.
\bibitem{experimental}
Devices were fabricated as in
~\onlinecite{barreiro09}, where a graphene sheet is mechanically exfoliated
from graphite onto a Si wafer, then structured using dry etching, and
finally electrically contacted to metal electrodes in a four-point
configuration.  
Raman spectroscopy and/or measurements of the reflected light intensity
with an optical microscopy are used to check that the device is a single layer.
The current is applied along the channel
while $V$ is measured between the voltage electrodes. 
Measurements are done at 300 K and 10$^{-6}$ mbar. 
$\alpha$ is obtained from the linear fit of the $I$-$V$
in a double logarithmic scale plot for $V>0.1$~V.
The mobility  is measured as $\mu=d\sigma/(d V_g)/C_g$,
being $\sigma(V_g)$ the zero-bias conductivity as a function of $V_g$.
$\mu$ is measured in the region where $\sigma(V_g)$ is linear
(which is usually attributed to the presence of Coulomb scatterers~\cite{chen08}).
\end{thebibliography}
\end{document}